\documentstyle[11pt,paspconf]{article}

\def\plotfiddle#1#2#3#4#5#6#7{\centering \leavevmode
\vbox to#2{\rule{0pt}{#2}}
\includegraphics{#1}}

\begin{document}

\title{Photometric Redshifts for DPOSS Galaxy Clusters at $z<0.4$}

\author{R. R. Gal, S. C. Odewahn, S. G. Djorgovski and R. Brunner}
\affil{Palomar Observatory, Caltech, Pasadena, CA 91106 }

\author{R. R. DeCarvalho}
\affil{Observatorio Nacional/CNPq, Brazil}




\begin{abstract}
We report on the creation of an unbiased catalog of galaxy clusters from the galaxy catalogs derived from the digitized POSS-II (DPOSS). Utilizing the g-r color information, we show that it is possible to estimate redshifts for galaxy clusters at z<0.4 with an rms accuracy of 0.01.
\end{abstract}




\section{Introduction}
There are many cosmological uses for rich clusters of galaxies.  They provide
useful constraints for theories of large-scale structure formation and
evolution, and represent valuable (possibly coeval) samples of galaxies to
study their evolution in dense environments. Studies of $\xi_{cc}$, the cluster two-point correlation function, are a powerful probe of
large-scale structure, and the scenarios of its formation. Until recently, it has been impractical to obtain large numbers of redshifts for glaxy clusters, forcing cosmologists to deproject their distribution mathematically. We show that it is feasible to generate a catalog of galaxy clusters at $ z<0.4$ with accurately estimated photometric redshifts.

\section{Observations}
Our data is taken from the Digitized Second Palomar Observatory Sky Survey (DPOSS). The digitization, star-galaxy separation, and photometric calibration procedures are described in Weir et al. (1995). We have improved star-galaxy classification using a much larger training set. 

We use a simple color selection of candidate cluster galaxies, coupled with the adaptive kernel method (Silverman, 1986) to generate galaxy surface density maps.  A bootstrap technique is then used to generate the statistical
significance map associated with a given surface density map. This map is then used to detect overdensities of galaxies on the sky which indicate candidate galaxy clusters.  In our test fields, we recover all 
of the known Abell clusters, and find a large number of new clusters.

%

\section{Redshift Estimation}
\begin{figure}[t]
\plotfiddle{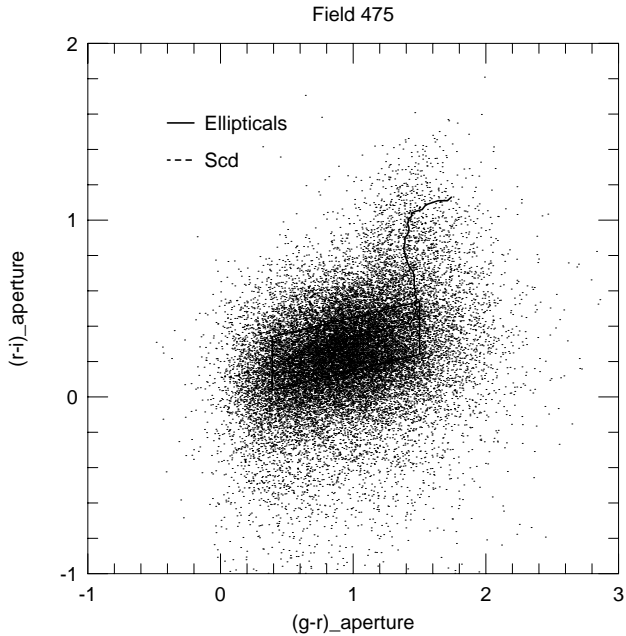}{2.6 in}{-90}{50}{50}{-190}{300}
\vskip -1.5 truecm
\caption {Color-color diagram for DPOSS glaxies} \label{fig-1}
\end{figure}

Because the $4000\AA$  break is shifting through the blue bandpass of
DPOSS at $z<0.4$, the $g-r$ color changes rapidly with redshift. We make the crude assumption that all cluster galaxies are a single--age, early--type population, and use a k--correction model to estimate redshifts from the $g-r$ color alone. We simply use the mean $g-r$ color of the 
galaxies in a cluster, after a background correction, to estimate the redshift.
In Figure~\ref{fig-1}, a $g-r$ vs. $r-i$ diagram for galaxies to $M_{r}=19.6$ in a typical DPOSS field (36 sq. deg.) is shown. Also shown are the k-correction tracks for
    Scd and E galaxies. The rapid change in $g-r$ between $z=0$ and 
    $z=0.4$ for early type galaxies allows us to estimate redshifts for
    galaxy clusters.

A separate redshift estimate, from the magnitude of the n-th brightest galaxy, can also be made, but it is much more sensitive to errors in the correction for field galaxies. 

\subsection{Technique}

In practice, the redshift estimation must be done iteratively. First, we detectcandidate clusters in our galaxy density maps. From those areas in our maps where there are no clusters, we estimate the background galaxy density and $g-r$ color distribution. This background correction is then applied to each cluster candidate in a fixed radius, corresponding to an Abell radius at $z=0.15$, the expected median redshift of our clusters. 

The redshift of each cluster is then estimated from the mean $g-r$ color of the galaxies inside this radius. Using this redshift estimate, we recalculate $R_{Abell}$, and estimate the redshift using the mean color in this radius. This procedure is repeated until the estimated redshift converges.

\begin{figure} [t]
\plotfiddle{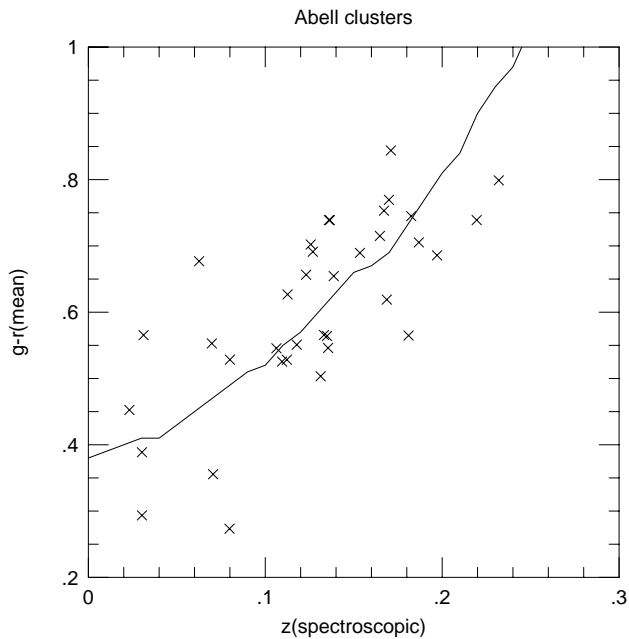}{4.2 in}{0}{50}{50}{-180}{-10}
\vskip -2.4 truecm
\caption {Redshift vs. $g-r$ color for 36 Abell clusters.} \label{fig-2}
\end{figure}

In Figure~\ref{fig-2}, we show the mean $g-r$ color for galaxies in 36 Abell 
    clusters with spectroscopic redshifts. The line shown is a theoretical
    k--correction track for E--type galaxies. It is NOT a fit to the data.
    The mean deviation of the data from the theoretical curve is $\Delta z=0.004$.
    This suggests that we can estimate redshifts for our candidate
    clusters in an accurate, unbiased way, directly from calibrated
    plate photometry. 

This result relies on the large number of same age and type galaxies in clusters at low redshift. As the cluster galaxy population changes with redshift, this technique will eventually fail. For $0.4<z<0.8$, the $r-i$ color could be used; unfortunately, our $i$ plate data are not deep enough for this purpose. We have obtained deeper CCD imaging of our low--$z$ candidates, where we will attempt to detect and estimate redshifts for more distant clusters.

\acknowledgments
R. Gal was supported in part by NASA GSRP NGT5-50215 and a Kingsley Fellowship. SGD acknowledges the support of the Norris Foundation. 


%
%

%


\begin{thebibliography}{}
\bibitem{}
Dressler, A. and Gunn, J.E., (1992) {\it Ap. J.}, {\bf Vol.~no.~78}, 1
\bibitem{}
Picard, A. 1992, Ph.D Thesis, Caltech
\bibitem{}
Silverman, B.W. 1986, {\it ~~Density ~~Estimation for Statistics and
Data Analysis}, (London: Chapman \& Hall)
\bibitem{}
Weir, N., Djorgovski, S., Fayyad, U., 1995 AJ, 110, 1
 
\end{thebibliography}
\end{document}